\documentclass[aps,prl]{revtex4}
\usepackage{amsmath} 
\usepackage{amssymb}    
\usepackage{graphicx} 

\begin{document}

\title{Universal route to thermalization in weakly-nonlinear one-dimensional  chains}

\author{
  L. Pistone$^1$}
  \author{
  S. Chibbaro$^2$}
    \author{
 M. D. Bustamante$^3$} 
 \author{
  Y. L'vov$^4$}
  \author{
  M. Onorato$^{1,5}$}


\affiliation{
$^1$ Dipartimento di Fisica, Universit\`a di Torino, via Pietro Giuria 1, 10125, Torino, Italy\\
$^2$ Sorbonne Universit\'e, CNRS, Institut Jean Le Rond D'Alembert, F-75005 Paris, France\\
$^3$School of Mathematics and Statistics, University College Dublin, Belfield, Dublin 4, Ireland\\
$^4$Department of Mathematical Sciences, Rensselaer Polytechnic Institute, Troy, NY 12180\\
$^5$ INFN, Sezione di Torino, Via P. Giuria, 1 - Torino, 10125, Italy\\
}


\begin{abstract} 
 We
      apply Wave Turbulence theory to describe
  the dynamics on nonlinear one-dimensional chains. We consider
      $\alpha$ and $\beta$ Fermi-Pasta-Ulam-Tsingou (FPUT) systems,
      and the discrete nonlinear Klein-Gordon chain. We demonstrate
  that resonances are responsible for the irreversible
  transfer of energy among the Fourier modes. We predict that all the systems thermalize for large
  times, and that the equipartition time scales as a power-law of the
  strength of the nonlinearity. Our methodology is not limited
      to only these systems and can be applied to the case of a
  finite number of modes, such as in the original FPUT experiment, or
  to the thermodynamic limit, i.e. when the number of modes approach
  infinity. In the latter limit, we perform  state of the art numerical simulations and show that the results are
  consistent with theoretical predictions. We suggest that the route to thermalization, based only on the
  presence of exact resonance, has universal features. Moreover, a by-product of our analysis is the 
asymptotic integrability, up to four wave interactions,  of the  discrete nonlinear Klein-Gordon chain.
\end{abstract}

\maketitle

\section{Introduction}

The numerical experiment of Fermi, Pasta, Ulam and Tsingou (FPUT) in 1955\cite{fermi1955studies} has been of great influence to physics in many respects. 
The idea of the experiment was related to the role of chaos on the foundations of statistical mechanics. 
As chaotic systems are highly irregular, like stochastic ones, they loose memory of initial conditions rapidly, as asked by ergodic hypothesis.
In this perspective, the result of Poincar\'e about the non existence  in generic Hamiltonian systems
of first integrals of motion, other than the energy, seems positive. 
In his early theoretical activity, Fermi generalised the result of  Poincar\'e showing that in Hamiltonian systems with $N>2$ degrees of freedom no smooth surface can divide the phase space into two regions containing invariant sets.
From this correct result, Fermi deduced that non-integrable hamiltonian systems are generically ergodic and, even in absence of a rigorous proof, the ergodic problem was considered basically solved. 
However, Fermi came back to the problem after the war with this numerical experiment, probably too much absorbed by the development of the just born quantum mechanics in the meanwhile.

With one of the first available computers, they investigated the
dynamics of a simple system of springs and masses, where the force is
only between nearest neighbors. The novelty consisted in the
metodology (it was one of the first numerical
studies), but also in the fact that the force was not linear.  In the
case of linear forces, it is known that the Hamiltonian of the system
can be diagonalized, the eigenstates are linear waves, and there is no
interaction between them, namely the initial energy distribution among
the modes does not change over time.  Fermi and collaborators wanted
to test whether introducing a small nonlinear term in the equations of
motion (they used quadratic, cubic and split-linear terms) would allow
energy to spread among the linear modes, and attain equipartition,
confirming the earlier idea of Fermi that ergodic hypothesis can be
considered true, giving a dynamical justification to statistical
mechanics. However, this was not the case and instead quasiperiodic
motion was observed or, at most, just a few modes close to the ones
initially excited were active for the whole simulation time.

The surprising results of FPUT generated a lot of  interest, and sparked a
research that resulted in a large body of work, as witnessed by an
excellent review\cite{gallavotti2008fermi}.  In particular, even
though ergodicity and chaos have been clarified to be largely
irrelevant for statistical mechanics of macroscopic systems, that is
when
$N\gg1$\cite{aleksandr1949mathematical,landau1980statistical,lebowitz1993boltzmann},
the existence of solitons arise from a possible explanation of the
FPUT phenomenology\cite{zabusky1965}.  Moreover, the FPUT experiment
was the first showing that numerical simulations are a powerful
instrument to get physical insights of complex phenomena, and
therefore an unavoidable tool for theoretical progress.  In parallel,
in the same time  it was shown by
mathematicians with the KAM
theorem~\cite{kolmogorov1954,arnold1963,moser1962} that generic
non-integrable Hamiltonian systems are not ergodic for small
perturbations. This fundamental result showed that the first guess by
Fermi was incorrect.

Despite these outstanding developments and considerable efforts, a clear and exhaustive explanation of the FPUT problem is still lacking.
In particular, while there is some consensus concerning the metastability picture~\cite{fucito1982approach,benettin2011time},
the precise mechanisms driving to equipartition on very long time-scales is still elusive.
One crucial point is that studies have focused on peculiar initial conditions and the specific phenomena that emerge from them in the short or medium timescales. For example, in the original experiments and in most of the successive studies the initial conditions consisted in exciting the low-frequency modes\cite{Benettin2009}, which leads to the formation of coherent nonlinear structures, that is solitons, at least in the short-time. 
Some attention has been focused also on high-frequency initial conditions and this led to the study of breathers.
 However, it is difficult to understand all these peculiar direction in one larger vision of the FPUT problem. 
The problem with generic initial conditions seems to be yet more relevant from a general perspective, as recognised in an important recent contribution\cite{carati2007fermi}.

In the present work, we propose a more unifying
picture; therefore, we seek universality traits that can explain the
dynamics in a way that is only weakly dependent on the particular
microscopic processes of the nonlinear chain in consideration. For
this reason, we will consider generic initial conditions that are
randomly out of equilibrium, involving low and high modes.
Moreover, we will use tools and methods of the so-called Wave
Turbulence (WT) theory. Originally introduced in fluid dynamics, WT is
a statistical mechanics theory of weakly interacting waves, and it has
been recently applied in relation to one dimensional anharmonic
chains.  In some sense, our approach is in line with the recent
results that have clearly showed that the $\alpha$-FPUT chain is a
perturbation of Toda lattice, which is an integrable
system\cite{benettin2013fermi}.  Our main motto is that the
irreversible transfer of energy in the spectrum in a weakly nonlinear
system is achieved by exact resonant wave-wave interactions. Such
resonances are the base for the WT theory and are responsible for the
phenomenon of thermalization.

In this paper we  give an overview on resonances and
WT in a comprehensive manner as well as present new 
results regarding the limit of a large
number of modes, providing insight in the
fundamental mechanics of anharmonic chains.  Our main result is
establishment of the power-law scaling of the equipartition time
$T_\text{eq}$ as a function of the nonlinearity strength (to be
defined precisely later). In varying the number of elements in the
chain, we demonstrate the
crossover between two different scaling laws (lower exponent for large
systems, that is the thermodynamic limit, steeper for smaller ones).

\section{The models}
We consider Hamiltonian systems of the form
\begin{equation}
H=H_{\text{lin}}+H_{\text{nlin}},
\end{equation}
where $H_{\text{lin}}$ is the integrable Hamiltonian, corresponding to a linear dispersive dynamics, and $H_{\text{nonlin}}$ that
contains the anharmonicity of the potential. More in particular we will deal 
with 
the $\alpha$, $\beta$-FPUT and the discrete nonlinear Klein-Gordon (DNKG) (also called the $\phi^4$ ) systems, all characterized by the linear Hamiltonian
\begin{equation}
H_{\text{lin}}=\sum_{j=0}^{N-1}\frac{1}{2}p_j^2+\frac{1}{2}(q_j-q_{j+1})^2+\frac{1}{2}mq_j^2,
\end{equation}
where $N$ is the number of elements in the chain, $q_j$ is the $j$-th coordinate and $p_j$ its conjugate momentum and $m$ models the linear part of the site-potential. For the $\alpha$ and $\beta$-FPUT systems $m=0$.
The nonlinear part of the Hamiltonian for the three models is given by:
\begin{equation}
H_{\text{nlin}}^{(\alpha)}=\frac{\alpha}{3}\sum_{j=0}^{N-1} (q_j-q_{j+1})^3,\;\;\;
H_{\text{nlin}}^{(\beta)}=\frac{\beta}{4}\sum_{j=0}^{N-1}(q_j-q_{j+1})^4,\;\;\;\;
H_{\text{nlin}}^{(KG)}=\frac{\beta}{4}\sum_{j=0}^{N-1}q_j^4.
\label{nonlinham}
\end{equation}
 Without loss of generality, we use the same parameter $\beta$ in front for nonlinearity in the DNKG and the $\beta$-FPUT.
We will use periodic boundary conditions. It should be noted, given an initial energy,
 that there is a finite yet very small probability that the $\alpha$-FPUT
chain breaks up, since
  the nonlinear part of the Hamiltonian is not bounded from
  below. While being an inherently ill-posed problem, it has been
  shown that for low enough energies the $\alpha$-FPUT chain can be
  stable for long times\cite{Carati2018}. For the other models, we
  restrict ourselves to the case where the quartic terms are positive,
  that is $\beta>0$, so that no blow-ups can occur. The parameter $m$
  is constrained to be greater than zero in the DNKG model, and we can
  anticipate that it will play a secondary role since we are
  interested in the effects of the nonlinear interactions. The
  presence of the quadratic finite difference terms in the linear
  Hamiltonian is important, because without that the eigenstates of
  the linear system would not be waves.

Our approach is of perturbative nature, hence
we need a way to quantify the nonlinearity strength. In general, the
linear and nonlinear parts of the Hamiltionan are not conserved
separately, but only the total Hamiltonian is. However, in typical
situations where the nonlinear interactions are not too strong,
operatively, we define the following nondimensional parameter for the
$\alpha$-FPUT
\begin{equation}
\label{eq:epsilon_a}
\epsilon_{\alpha}= \left(\frac{\sum_{j=0}^{N-1}\left|h_\text{nlin}^{(\alpha)}(j)\right|}{H_{\text{lin}}}\right)^{2}\propto \alpha^{2},
\end{equation}
and for the $\beta$-FPUT and DNKG
\begin{equation}
\label{eq:epsilon_b}
\epsilon_{\beta,\text{KG}}= \frac{H_{\text{nlin}}^{(\beta,\text{KG})}}{H_{\text{lin}}}\propto \beta.
\end{equation}
 $h_\text{nlin}^{(\alpha)}(j)$ is the nonlinear part of the Hamiltonian density of  the $j$-th particle  in the for $\alpha$-FPUT. The absolute value in eq.~(\ref{eq:epsilon_a}) is necessary  because $h_\text{nlin}(j)$ can be negative and cancellations in the sum would cause an incorrect accounting of the nonlinearity strength. The different scalings in equations (\ref{eq:epsilon_a}) and (\ref{eq:epsilon_b}) are due to the different degrees of nonlinearity between the  $\alpha$-FPUT and the other two models. Definitions~(\ref{eq:epsilon_a}) and (\ref{eq:epsilon_b}) allows us to refine our statement that the nonlinearity must be small, that is, we require that the parameters are much less than one. In general, they fluctuate over time, so an appropriate averaging is needed to have a meaningful estimate of the nonlinearity strength. The parameters are further discussed when presenting the numerical results.

\subsection{Fourier space}

The eigenstates of the linear Hamiltonian are waves, hence it is useful to work in Fourier space. We define the direct and inverse discrete Fourier transform of the $q_j$ variables,
\begin{equation}
\label{eq:fourier}
\hat q_k=\frac{1}{N}\sum_{j=0}^{N-1} q_je^{-i2\pi kj/N},\ q_j=\sum_{k=0}^{N-1}\hat q_ke^{i2 \pi kj/N}
\end{equation}
and similar definitions  for
$\hat p_k$. We will use the convention that $0\leq k< N$, and we note
$\hat q^*_k=\hat q_{N-k}$ and $\hat p^*_k=\hat p_{N-k}$.  After this
change of variables, and using $\sum_{j=0}^{N-1} e^{i2\pi
  (k_1-k_2)j/N}=N\delta_{k_1-k_2}$ with
$\delta^{(N)}_{k_1+k_2+...}=\delta(k_1+k_2+... \mod N)$ that is the
Kronecker's delta modulo $N$, the linear part of the Hamiltonians can
be written as
\begin{equation}
\label{eq:Hlinfourier}
\frac{H_{\text{lin}}}{N}=\frac{1}{2}\sum_{k=0}^{N-1}\left(\hat p_k^2+\omega_k^2\left|\hat q_k^2\right|\right),
\end{equation}
where we have defined the linear dispersion relation
\begin{equation}
\label{eq:omega}
\omega_k=\sqrt{m+4\sin^2\left(\pi\frac{k}{N}\right)}
\end{equation}
(we recall that $m=0$ for the FPUT models).  The nonlinear part of the
Hamiltonians couple the modes and can be
interpreted as n-wave collision terms.  For the
sake of brevity, we denote $\hat q_{j}= \hat q_{k_j}$ and
$\delta^{(N)}_{1+2+...}=\delta(k_1+k_2+... \mod N)$.  For the FPUT
models we obtain
\begin{equation}
\frac{H_{\text{nlin}}^{(\alpha)}}{N}=\frac{\alpha}{3}\sum_{k_1,k_2,k_3=0}^{N-1}\widetilde{A}_{1,2,3}\hat q_{1}\hat q_2\hat q_3\delta^{(N)}_{1+2+3},\;\;\;
\frac{H_{\text{nlin}}^{(\beta)}}{N}=\frac{\beta}{4}\sum_{k_1,k_2,k_3,k_4=0}^{N-1}\widetilde{B}_{1,2,3,4}\hat q_{1}\hat q_2\hat q_3\hat q_4\delta^{(N)}_{1+2+3+4},
\end{equation}
where the collision matrices $\widetilde{A}_{1,2,3}$ and $\widetilde{B}_{1,2,3}$ are (after symmetrization)
\begin{equation}
\widetilde{A}_{1,2,3}=8ie^{i\pi(k_1+k_2+k_3)/N}\sin\left(\pi \frac{k_1}{N}\right)\sin\left(\pi \frac{k_2}{N}\right)\sin\left(\pi \frac{k_3}{N}\right),
\end{equation}
\begin{equation}
\widetilde{B}_{1,2,3,4}=16e^{i\pi(k_1+k_2+k_3+k_4)/N}\sin\left(\pi \frac{k_1}{N}\right)\sin\left(\pi \frac{k_2}{N}\right)\sin\left(\pi \frac{k_3}{N}\right)\sin\left(\pi \frac{k_4}{N}\right).
\end{equation}
The complex exponential is relevant only when crossing Brillouin zones in the sum of wave numbers. For the DNKG model, it turns out that the interaction matrix is equal to unity, hence the nonlinear part of the Hamiltonian is simply
\begin{equation}
\frac{H_{\text{nlin}}^{(KG)}}{N}=\frac{\beta}{4}\sum_{k_1,k_2,k_3,k_4=0}^{N-1}\hat q_{1}\hat q_2\hat q_3\hat q_4\delta^{(N)}_{1+2+3+4}.
\end{equation}

From eq.~(\ref{eq:Hlinfourier}), we can further simplify the problem
using the normal modes representation,
\begin{equation}
a_k=\frac{1}{\sqrt{2\omega_k}}\left(\hat p_k - i \omega_k\hat q_k \right),\ a^*_{N-k}=\frac{1}{\sqrt{2\omega_k}}\left(\hat p_k + i \omega_k\hat q_k\right)
\end{equation}
as the linear part of the Hamiltonians becomes simply
\begin{equation}
\label{eq:Hlin}
\frac{H_{\text{lin}}}{N}=\sum_{j=0}^{N-1}\omega_k\left|a_k\right|^2,
\end{equation}
whereas the nonlinear terms in the FPUT models and in the DNKG become (after renaming $N-k_i=k'_i$ where needed and dropping the prime)
\begin{equation}
\label{eq:Hnlin}
\frac{H_{\text{nlin}}^{(\alpha)}}{N}
=\alpha\sum_{k_1,k_2,k_3=0}^{N-1}A_{1,2,3}\left[\frac{1}{3}(a_1a_2a_3+c.c.)\delta^{(N)}_{1+2+3}+(a^*_1a_2a_3+c.c.)\delta^{(N)}_{1-2-3}\right],
\end{equation}
\begin{equation}
\begin{split}
\frac{H_{\text{nlin}}^{(\beta,\text{KG})}}{N}=\beta\sum_{k_1,k_2,k_3,k_4=0}^{N-1}B_{1,2,3,4}^{(\beta,\text{KG})}\left[\frac{1}{4}(a_1a_2a_3a_4+c.c.)\delta^{(N)}_{1+2+3+4}+(a^*_1a_2a_3a_4+c.c.)\delta^{(N)}_{1-2-3-4}+\right.\\
\left.+\frac{3}{2}a^*_1a^*_2a_3a_4\delta^{(N)}_{1+2-3-4}\right],
\end{split}
\end{equation}
with the interaction matrices given by
\begin{equation}
\label{eq:AFPUT}
A_{1,2,3}=\frac{\widetilde{A}_{1,2,3}}{2^{3/2}\sqrt{\omega_1\omega_2\omega_3}},\;\;\;
B_{1,2,3,4}^{(\beta)}=\frac{\widetilde{B}_{1,2,3,4}}{4\sqrt{\omega_1\omega_2\omega_3\omega_4}},\;\;\;
B_{1,2,3,4}^{\text{(KG)}}=\frac{1}{4\sqrt{\omega_1\omega_2\omega_3\omega_4}}.
\end{equation}
The nonlinear terms  show that a nonlinearity in physical space turns into n-mode collision terms such as $a_1a_2a_3\delta^{(N)}_{1+2-3}$. 
From the Hamilton equations, we can obtain the dynamical equations $\dot{a}_k=-i(1/N)\delta H/\delta a_k^*$, which read for the 
$\alpha$-FPUT
\begin{equation}
\label{eq:eqm_a}
i\dot{a}_1=\omega_1 a_1+\alpha\sum_{k_1,k_2,k_3=0}^{N-1}A_{1,2,3}\left(a_2a_3\delta^{(N)}_{1-2-3}+2a^*_2a_3\delta^{(N)}_{1+2-3}+a^*_2a^*_3\delta^{(N)}_{1+2+3}\right)
\end{equation}
and for the $\beta$-FPUT and DNKG
\begin{equation}
\label{eq:eqm_b}
\begin{split}
i\dot{a}_1=\omega_1 a_1
+\beta\sum_{k_1, k_2,k_3,k_4=0}^{N-1}B_{1,2,3,4}^{(\beta,\text{KG})}\big(&a_2a_3a_4\delta^{(N)}_{1-2-3-4}+3a^*_2a_3a_4\delta^{(N)}_{1+2-3-4}+\\
&+3a^*_2a^*_3a_4\delta^{(N)}_{1+2+3-4}+a^*_2a^*_3a^*_4\delta^{(N)}_{1+2+3+4}\big).
\end{split}
\end{equation}

These types of Hamiltonians are the canonical form of Hamiltonians
    for system composed of interacting particles or waves. If nonlinearity
    is small in the above mentioned sense, then there is a well developed
    theory called Wave Turbulence theory, which is described in details in
\cite{nazarenko2011wave,falkovich1992kolmogorov} and briefly in the subsequent section.

The number of modes in the nonlinear terms defines the collision
order, that is the $\alpha$-FPUT contains 3-wave collisions, while
$\beta$-FPUT DNKG 4-wave collisions (we will see that the effective
collision order in the $\alpha$-FPUT is also of the 4-wave type).  The
number of non-conjugate and conjugate variables (which matches the
number of positive vs. negative terms in the Kronecker's deltas)
defines different collision processes,
e.g. $a_1a_2a_3\delta^{(N)}_{1+2+3}$ is a $3\rightarrow 0$ collision
process (annihilation), while $a_1a_2a^*_3a^*_4\delta^{(N)}_{1+2-3-4}$
is a $2\rightarrow 2$ collision process (scattering).

The WT approach starts from the equations of motion in the canonical variables $a_k$. 
Note that so far we have not taken any approximation; we only made the choice of using periodic boundary conditions.

\section{Wave-Turbulence theory in a nutshell}
When the number of modes is large, the microscopic dynamics given by
equations (\ref{eq:eqm_a}) and (\ref{eq:eqm_b}) do not provide much analytical insight and a
statistical approach is needed.  WT is the general statistical theory
valid in the weak-nonlinear regime.  While the theory has been
recently developed for higher-order statistical
observables\cite{nazarenko2011wave,eyink2012kinetic,chibbaro20184} ,
the core of the WT is the kinetic equation developed for the
prediction of the energy spectrum.  The basic statistical observable
is the two point correlator $ \langle a_1a_2^*\rangle$ which, under
the hypothesis of homogeneity, it is given by
\begin{equation}
\label{eq:n}
\langle a_1a_2^* \rangle =n_{1} \delta(k_1-k_2),
\end{equation}
with $n_k=n(k,t)$ the wave actions spectral density, i.e. the Fourier
transform of the autocorrelation function of $a(x,t)$.  The average
is over an ensemble of realizations
of the system with the same initial linear energy $H_\text{lin}$.
Then, the energy spectrum that gives the mean energy per mode is
\begin{equation}
\label{eq:e}
e_k=\omega_kn_k.
\end{equation}

The evolution equation for the wave action spectral density can be
obtained by various
    techniques~\cite{falkovich1992kolmogorov,newell2011wave,nazarenko2011wave,eyink2012kinetic,chibbaro20184}.
The main issue is due to the fact that calculating the evolution of
$n_k$ from the equation of motion~(\ref{eq:eqm_a}) or
(\ref{eq:eqm_b}), one encounters the well know problem of the BBGKY
hierarchy\cite{during2017wave}, that is, the evolution of lower order
correlators depends on higher order correlators and a closure problem
arises.  Originally, a closure was obtained with a quasi-gaussian
approximation which allowed to expresses higher cumulants in terms of
lower ones\cite{falkovich1992kolmogorov}.  It
    turns out that less stringent hypotheses are necessary. It is sufficient to
make a random-phase approximation and to consider initial conditions
which have also random
amplitudes~\cite{nazarenko2011wave,eyink2012kinetic,chibbaro20184}.
The random phase assumption is generally deemed to be solid because
the linear waves decorrelate quickly even in the linear regime, hence
it is expected that such property still holds true in the weakly
nonlinear regime, as corroborated by numerical
experiments~\cite{chibbaro2017wave} In all the following treatment, we
assume that this property holds, and that the phases are independent
random variables uniformly distributed between $[0,2\pi)$.  The
  resulting kinetic equation can be obtained in the thermodynamic
  limit $N\rightarrow \infty$.  Generally also the continuum limit is
  taken at the same time, and both physical ${\bf x}$ and momentum
  space ${\bf k}$ are continuous.

The main concept underneath the kinetic equation is the existence of
conservation laws associated to the wave scattering processes. Indeed,
each nonlinear term in the equations (\ref{eq:eqm_a}) and
(\ref{eq:eqm_b}) contains an appropriate Kroneker $\delta$
function over wave numbers. Depending on the shape of the
dispersion relation, it may be possible to associate also a related
conservation of energy. To be more specific, let us consider for
example the scattering processes,
\begin{equation}
\label{eq:momentum}
k_1=k_2+k_3,
\end{equation}
 contained  in equation  (\ref{eq:eqm_a}), then the main question is to verify if a similar relation holds for frequencies, i.e.:
 \begin{equation}
 \label{eq:energy}
\omega_1=\omega_2+\omega_3.
\end{equation} 
Equations (\ref{eq:momentum}) and (\ref{eq:energy}) define a resonant
manifold in wave number space, whose existence, as anticipated,
depends only the analytical form of the dispersion curve
$\omega_k$. If equation (\ref{eq:momentum}) and (\ref{eq:energy}) are
satisfied for the same wave numbers, then we are dealing with a {\it
  resonant} process which, according to the kinetic equation, will
lead to an irreversible transfer of energy.

There are two main types of wave systems that are
    typically considered in the framework of WT: the one dominated by three wave resonances
    and by four wave resonances (higher order are also possible). Examples of the systems dominated by
    three wave resonances include capillary waves on a surface of
    fluid and internal waves in the ocean.  Examples of four wave resonant
    systems include gravity waves on deep water, as well as the celebrated 2D-Nonlinear
Schr\"odinger equation. 

For capillary waves on a surface  three wave resonant interactions are possible  and a kinetic equation can be derived, \cite{falkovich1992kolmogorov}:
 \begin{multline}
\label{eq.kinetic-a}
\frac{\partial n_1}{\partial t}=\int_{-\infty}^{+\infty}A_{123}^2n_{1}n_{2}n_{3}\left(\frac{1}{n_{1}}-\frac{1}{n_{2}}-\frac{1}{n_{3}}\right)\delta({\bf k}_1- {\bf k}_2- {\bf k}_3)\delta(\omega_1- \omega_2- \omega_3)  {\mathrm{d}}{\bf k}_{23}+2 \left \{({\bf k}_1 \leftrightarrow {\bf k}_2)\right \},
\end{multline}
where $ {\bf k}$ is a two dimensional wave vector,  ${\mathrm{d}}{\bf k}_{23}= {\mathrm{d}}{\bf k}_{2} {\mathrm{d}}{\bf k}_{3}$ and $\left \{({\bf k}_1 \leftrightarrow {\bf k}_2)\right \}$ implies the same integral with ${\bf k}_1$ and $ {\bf k}_2$ exchanged. $A_{123}$ is a matrix that can be derived directly from the dynamical equations  for capillary waves.

 For surface gravity waves in deep water, it can be show that the leading order resonant process is the four-wave one, characterized by the following resonant process:
\begin{equation}
\label{eq:resonance}
\begin{split}
&{\bf k}_1+{\bf k}_2={\bf k}_3+{\bf k}_4 \\
&\omega_1+\omega_2=\omega_3+\omega_4.
\end{split}
\end{equation} 
The appropriate kinetic equation describing such process is 
 \begin{multline}
\label{eq.kinetic-b}
\frac{\partial n_1}{\partial t}=\int_{-\infty}^{+\infty}B_{123}^2n_{1}n_{2}n_{3}n_{4}\left(\frac{1}{n_{1}}+\frac{1}{n_{2}}-\frac{1}{n_{3}}-\frac{1}{n_{4}}\right)\delta({\bf k}_1+ {\bf k}_2- {\bf k}_3- {\bf k}_4)\delta(\omega_1+\omega_2- \omega_3- \omega_4)  {\mathrm{d}}{\bf k}_{234}.
\end{multline}
Note that in both kinetic equation the interaction matrix is squared. 
The latter equation, with a transport term, a forcing term due to the wind and term mimicking the dissipation, is integrated daily for operationally ocean wave forecasting pourposes, see \cite{janssenbook04}.

The kinetic equations~(\ref{eq.kinetic-a}) and (\ref{eq.kinetic-b}) have a number of important properties. First of all, it can be shown that both equations conserves the total energy and momentum:
\begin{equation}
\label{eq:E}
E=\int\omega_kn_{\bf k}\mathrm{d}{\bf k},\;\;\;\;\;\;\;\; {\bf M}=\int {\bf k} n_{\bf k} \mathrm{d}{\bf k}.
\end{equation}
The kinetic equation for four-wave interactions preserves also the number of waves (particles) in the scattering:
\begin{equation}
\label{eq:N}
\mathcal{N}=\int n_{\bf k}\mathrm{d}{\bf k}.
\end{equation}
The kinetic equation~(\ref{eq.kinetic-b}) is reminiscent of the Boltzmann equation for hard spheres\cite{Spohn2006} and there exists an entropy function
\begin{equation}
\label{eq:entropy}
S=\int\mathrm{d}{\bf k}\log(n_{\bf k})
\end{equation}
such that $\mathrm{d}S/\mathrm{d}t\geq0$ (the same entropy can be defined for the three wave kinetic equation). The isotropic equilibrium is reached when  $\mathrm{d}S/\mathrm{d}t=0$, i.e. at  the Rayleigh-Jeans distribution:
\begin{equation}
\label{eq:RJ}
\left.n_k\right|_{\text{equilibrium}}=\frac{T}{\omega_k+\mu},
\end{equation}
where $T$ is some normalization constant linked to the total energy and $\mu$ is a chemical potential linked to the conservation of the total number of particles~(\ref{eq:N}) and it is zero in the case of a three wave process. We see that when $\mu=0$ then WT predicts a relaxation towards equipartition of energy.
In this framework, exact resonances  bring the system to thermalization with a time-scale which is proportional to the coupling coefficient in front of the collision integral.

Some remarks are in order. 
\begin{itemize}
\item Given a physical dispersive waves system, it has to be checked if exact resonances are allowed.
In some cases, only trivial collisions or processes which are not able to transfer energy are found. In those cases, it is necessary to pursue the perturbative procedure and derive a kinetic equation with higher-order processes to check if at some point resonances are found. 
This issue is related to the dispersion relation, and in physical phenomena $3-$, $4-$, $5-$ and $6-$wave processes have been recognised\cite{nazarenko2011wave}.
In other cases, it can be shown that no exact resonances exist and therefore no exchange of energy is possible,
as rigorously shown to be the case for integrable systems \cite{zakharov1988additional}.
Other systems may show some resonances which are not yet able to transfer energy among modes because resonating waves belong to isolated clusters \cite{bustamante2011resonance}.
\item The WT framework is an asymptotic theory valid  in the weakly nonlinear r\'egime. If the nonlinearity is not small, different effects can play a role. In particular, the broadening of the frequencies due to the finite value of the nonlinear coupling, may trigger some resonances which were not present in the weak asymptotic limit, therefore allowing a more efficient transfer of energy.
\end{itemize}

In the present paper, we estimate the
time scale of equipartition; therefore, it is of major importance to
be able to derive a kinetic equation for our chain models. One should
note that chains are intrinsically discrete in physical space and, for
a finite number of masses, also the Fourier space is discrete. This
issue makes the derivation of the kinetic equation not straightforward
and, only in the thermodynamic limit, a formal derivation of the
kinetic equation is possible.  We shall deal with these points in the
following section in some detail.

\section{Resonance and  effective Hamiltonians in the thermodynamic limit}
We consider the thermodynamic limit which consists in taking the number of 
particles going to infinity, $N\rightarrow \infty,$ keeping constant the mass linear density. The general linear dispersion relation 
becomes 
\begin{equation}
\label{eq:omegacont}
\omega_{k}=\sqrt{m+4\sin^2\left(\frac{k}{2}\right)}
\end{equation}
with $k$ real in the interval  $0\le k<2\pi$.
We consider interaction processes from  $X$ to $Y$ waves for which 
 the following equations are satisfied:
\begin{equation}
\begin{split}
&k_1+k_2+...+k_X=k_{X+1}+k_{X+2}...+k_{X+Y}\;\;\; \mod 2\pi,\\
&\omega_1+\omega_2+...+\omega_X=\omega_{X+1}+\omega_{X+2}...+\omega_{X+Y}.
\end{split}
\end{equation}
Our goal is to find resonances. Once the leading 
order resonance (lowest value of $X$ and $Y$) has been found, using tools from Hamiltonian mechanics, an effective Hamiltonian
can be established and a kinetic equation can be built.

\begin{figure}
\centering 
\includegraphics[width=0.5\textwidth]{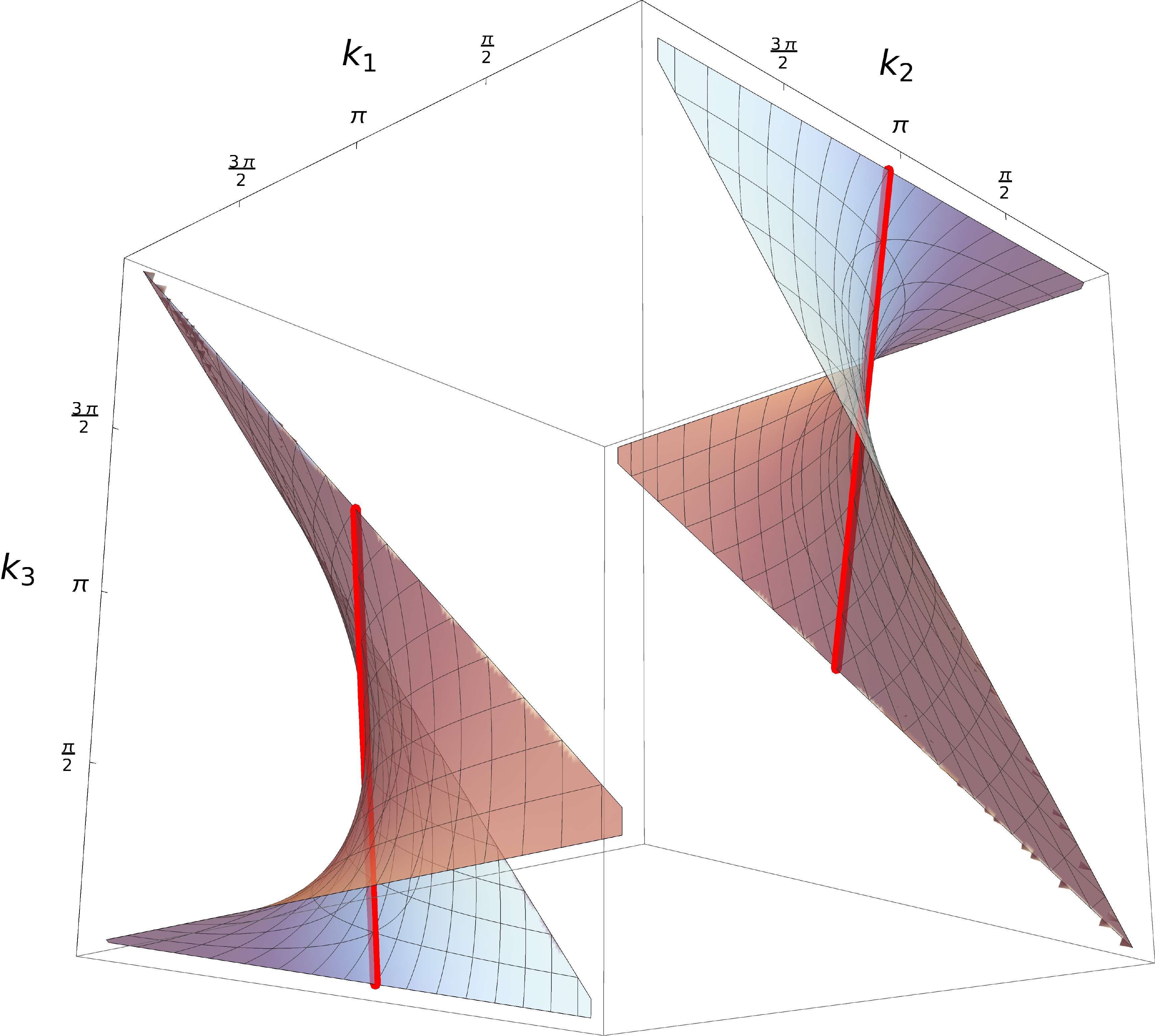}
\caption{The continuous resonant manifold for $2\rightarrow 2$ resonances with $m=0$, that is $\omega_1+\omega_2=\omega_1+\omega_{1+2-3}$. The trivial solutions $k_1=k_3$ or $k_2=k_3$ are not shown for clarity. The pairing-off resonances, to be discussed later, of eq.~(\ref{eq:umklapp}) are highlighted with the red line.}
\label{fig:manifoldcont}
\end{figure}

The first trivial observation is that resonances of the type $X$ to 0  are excluded for all three models; this is because $\omega_k\geq0$, and it is equal to zero only for $m=0$ for mode $k=0$ which does not play any role in the dynamics (for $m=0$). 
\subsection{Three-wave resonant interactions in the $\alpha$-FPUT}
The leading order nonlinearity in the evolution equation for the $\alpha$-FPUT is cubic; this implies three-wave interactions
of the form 2 $\rightarrow$ 1, 1 $\rightarrow$ 2 or 3 $\rightarrow$ 0 (the latter have been already excluded).
It is known, see for example \cite{matkowski2011subadditive,bustamante2018exact},  that the function $f_k=2|\sin(k/2)|$ is a subadditive function, i.e.
\begin{equation}
|\sin(k_1/2)|+|\sin(k_2/2)|\ge |\sin((k_1+k_2)/2)|.
\end{equation}
The equality holds only for $k_1$ or $k_2=l 2\pi$, with $l \in \mathbb{Z}$; those wave numbers do not enter into the dynamics. The subadditivity implies the non existence of three-wave resonant interactions for $m=0$.

\subsection{Four-wave resonant interactions in the $\alpha$, $\beta$-FPUT and DNKG models}
The  $\beta$-FPUT and DNKG models include processes involving four waves,
while, at first sight, the $\alpha$-FPUT model does not include them; however, as it will be shown 
in the following, the absence of three-wave resonances allows for a perturbative approach on the $\alpha$-FPUT model  that leads to four and higher order wave-wave interaction processes.
In \cite{bustamante2018exact} it has been shown that resonant processes 3 $\rightarrow$ 1 or 1 $\rightarrow$ 3 are excluded for the FPUT models, i.e. in the case $m$=0. In the Appendix  we show that this implies that also for the case of $m\ne0$ there are no such processes.   It turns out that in the continuous limit there exist, apart from trivial resonances
$k_1=k_2=k_3=k_4$ or $k_1=k_3$, $k_2=k_4$, a nontirvial  resonant manifold for interactions of the type $2\rightarrow 2$ (see figure~\ref{fig:manifoldcont} for a visualization of the case $m=0$, similar result apply for $m>0$). The result has been obtained using the software Mathematica Wolfram.

\subsection{The effective Hamiltonian and the kinetic equation for $\alpha$, $\beta$-FPUT and DNKG models}
The effective Hamiltonian is obtained by removing all non resonant interactions from the original one.
As standard in Hamiltonian systems, this can be carried out through a canonical change of variables (quasi-identity transformation), from the normal modes $a_k$ to some other variables $b_k$, such that in the new variables the nonresonant terms are removed from the Hamiltonian, generating higher order nonlinearities.  It is worth underlying that one has to ensure the canonicity of the transformation, at least up to the order of the new interaction terms that appear in the new variables. 
For example, to remove all the three-wave terms in the $\alpha$-FPUT model one uses
\begin{equation}
\label{eq:change}
a_1=b_1+\int_0^{2\pi} \left(X^{(1)}_{1,2,3}a_2a_3\delta^{(2\pi)}_{1-2-3}+X^{(2)}_{1,2,3}a^*_2a_3\delta^{(2\pi)}_{1+2-3}+X^{(3)}_{1,2,3}a^*_2a^*_3\delta^{(2\pi)}_{1+2+3}\right) {\mathrm{d}}k_2  {\mathrm{d}}k_3 +...,
\end{equation}
where the  integral  corresponds to three integrals from 0 to ${2\pi}$ and not from $-\infty$ to $+\infty$ 
and $\delta_{2\pi}$ account for periodicity of the Fourier space; this is because of the discreteness of our system in physical space.
In eq.~(\ref{eq:change}) we recognize terms equivalent to the nonlinear interactions in equation (\ref{eq:eqm_a}). The matrices $X^{(i)}_{1,2,3}$ are determined by inserting eq.~(\ref{eq:change}) into the $\alpha$-FPUT Hamiltonian and by grouping the terms corresponding to different wave processes. The transformation generates four-wave interactions (and higher) and again one has to look for resonances and remove the non resonant ones.
The calculation leads to the following effective Hamiltonian for the $\alpha$-FPUT
\begin{equation}
H^{(\alpha)}=\int_0^{2 \pi} \omega_k\left|b_k\right|^2{\mathrm{d}}k+
\frac{\alpha^2}{2}\int_0^{2\pi}\bar B_{1,2,3,4}^{(\alpha)}
b^*_1b_2^*b_3b_4\delta^{(2\pi)}_{1+2-3-4} {\mathrm{d}}k_2  
{\mathrm{d}}k_3{\mathrm{d}}k_4
\end{equation}

For the $\beta$-FPUT and for the DNKG a transformation is used to remove the 3$\rightarrow$1, 1$\rightarrow$3 and 
4$\rightarrow$0 terms, so that the effective Hamiltonian takes the form:
\begin{equation}
H^{(\beta,KG)}=\int_0^{2 \pi} \omega_k\left|b_k\right|^2{\mathrm{d}}k+
\frac{\beta}{2}\int_0^{2\pi}\bar B_{1,2,3,4}^{(\beta,KG)}
b^*_1b_2^*b_3b_4\delta^{(2\pi)}_{1+2-3-4} {\mathrm{d}}k_2  
{\mathrm{d}}k_3{\mathrm{d}}k_4.
\end{equation}
For an interested reader, a comprehensive procedure for removing three
and four-wave interactions, is presented in
\cite{laurie2012one,krasitskii1994reduced,Dyachenko1995233}.
All three models are dynamically described by the Zakharov
equation:
\begin{equation}
\label{eq:zakharov}
i\frac{\partial b_1}{\partial t}=\omega_1 b_1
+\int_0^{2\pi}W_{1,2,3,4}b^*_2b_3b_4\delta^{(2\pi)}_{1+2-3-4} {\mathrm{d}}k_2{\mathrm{d}}k_3{\mathrm{d}}k_4,
\end{equation}
where the coefficient $W_{1,2,3,4}$ takes different form, depending on
the particular system under consideration. Its
actual analytical form is irrelevant in our discussion (except in those cases for 
which the coefficient is zero on the resonant manifold \cite{zakharov1991integrability,zakharov2012integrable}); it is just sufficient to mention that it
is proportional to $\alpha^2 (A_{1,2,3})^2$ for the $\alpha$-FPUT
model and to $\beta B_{1,2,3,4}^{(\beta,KG)} $ for the $\beta$-FPUT
and DNKG models.

The Zakharov equation is the starting point for developing the statistical theory, i.e. the wave kinetic equation, which takes the following form (see
for details \cite{falkovich1992kolmogorov,nazarenko2011wave,chibbaro20184}):
\begin{equation}
\label{eq:kinetic4}
\frac{\partial n_1}{\partial t}=\int_0^{2\pi}\mathrm{d}k_{2,3,4}W_{1,2,3,4}^2\delta^{(2\pi)}(k_1+k_2-k_3-k_4)
\delta(\omega_1+\omega_2- \omega_3- \omega_4)n_1n_2n_3n_4\left(\frac{1}{n_1}+\frac{1}{n_2}-\frac{1}{n_3}-\frac{1}{n_4}\right).
\end{equation}
Note that, because of the $\delta^{(2\pi)}$, instead of $\delta$, the
momentum is not conserved.  The time scales, $T_\text{eq}$, of four
wave resonant interaction, which lead to a thermalized state,
are:
\begin{equation}
T_\text{eq}\propto\alpha^{-4}\propto\epsilon_{\alpha}^{-2}
\end{equation}
for the $\alpha$-FPUT model and
\begin{equation}
T_\text{eq}\propto\beta^{-2}\propto \epsilon_{\beta,KG}^{-2}
\end{equation}
for the DNKG and $\beta$-FPUT models.

We  we stress that
the theory developed in this section is rigorously valid when the
thermodynamic limit and the weak non-linear limit are taken, in the
given order. It is important to wonder if the scaling laws obtained
can be observed in actual finite-size systems and in particular in
numerical experiments. We expect a positive answer, at least for
sufficient large $N$ and small nonlinearity.  Indeed, it is well known
that a small nonlinearity cause a frequency shift of the linear modes,
and also a stocasticization of the frequencies, or in other words, a
broadening\cite{izrailev1966statistical,CHIRIKOV1979263}.  It is
reasonable to assume that resonances do not need to be satisfied
exactly in practical applications\cite{kartashova2007exact}.  It is
sufficient that broadening of frequencies becomes comparable with the
spacing of the frequencies, which decreases with the number of modes,
so that $\omega_k$ becomes continuous in the thermodynamic limit.  In
this sense, the discrete representation should converge to the
continuous theory, and the higher the number of modes the smaller the
nonlinearity required to be in agreement with the theory. We shall
check this argument with extensive numerical simulations in the
following.


\section{Discrete exact resonances}
In many interesting cases, the number of modes $N$ is not too large, like for instance in the original FPU numerical experiments.
In this case, it is not possible to consider the system a good approximation of the continuous one, and it has to be studied in its discrete form. As mentioned, one should recall that in discrete systems the condition in the Kroneker $\delta$ over wave numbers should be intended mod $N$.

\subsection{Four-wave resonant interactions, effective Hamiltonian and itegrability}
The first problem is to find out if discrete exact resonances exist and at which order. 
The previous discussion on three-wave resonance still applies here, hence three-wave resonances are always excluded from 
the $\alpha$-FPUT model.
Concerning 4-wave resonances, the only process that is potentially active for the dispersion relation~(\ref{eq:omega}) is the scattering process $2\rightarrow 2$. These resonances have been considered extensively in\cite{Onorato2015,pistone2018thermalization,bustamante2018exact}, and we will recap here the results. 
Obviously, even discrete processes of the type $X\rightarrow X$ admit trivial resonances of the type
\begin{equation}
\label{eq:trivial}
k_1+k_2=k_1+k_2\mod N,\,\omega_1+\omega_2=\omega_1+\omega_2
\end{equation}
or $k_1=k_2=k_3=k_4$.
These processes, however, do not result into an exchange of energy, but rather cause a frequency shift of the linear modes\cite{nazarenko2011wave}. Because the system is periodic and the dispersion relation is symmetric,
\begin{equation}
\label{eq:pair-off_omega}
\omega_{k}=\omega_{N-k}=\omega_{-k},
\end{equation}
it turns out that nontrivial resonances of the type $2\rightarrow 2$ are possible with the crossing of Brillouin zones
(Umklapp scatterings). For $N$ even, these resonances take the form
\begin{equation}
\label{eq:umklapp}
k_1+k_2=-k_1-k_2\mod N,\qquad k_1+k_2=N/2,
\end{equation}
and are known as   \textit{pairing-off} resonances \cite{bustamante2018exact}. However, one can easily check that these resonances are all disconnected, and they actually give rise to integrable dynamics \cite{henrici2008results,rink2006proof}. In\cite{pistone2018thermalization} other special resonances have been considered for some very specific values of $m\ne 0$, but they are limited in number and they do not appear in general to be able to cover the whole Fourier space (a detailed study should be performed for such specific cases).

The effective integrable Hamiltonian, up to four-wave interactions for the $\alpha$, $\beta$ and DNKG equations (with $m$ different for those special values for which other resonant quartets exists), can be recast as follows:
\begin{equation}
\begin{split}
\frac{H_{\text{integrable}}}{N}=&\sum_{k} \omega_{k} |b_{k}|^2 
+\frac{1}{2} \sum_{k} W_{k,k,k,k} \left(|b_{k}|^2\right)^2+  \sum_{k_1\ne k_2} W_{k_1,k_2,k_1,k_2} |b_{k_1}|^2|b_{k_2}|^2+\\
&
+\sum_{k=1}^{\lfloor N/4\rfloor}2 W_{k,\frac{N}{2}-k,-k,-\frac{N}{2}+k}\left(b_{k}^*b_{\frac{N}{2}-k}^*b_{-k}b_{-\frac{N}{2}+k}+c.c.\right),\end{split}
\end{equation}
where $c.c.$ denotes complex conjugate.
The first three terms depend on the moduli of the amplitudes only and underline an integrable dynamics (Birkhoff normal form); 
interestingly, the last term does not break integrability (resonant Birkhoff normal form)  \cite{henrici2008results,rink2006proof}, due to the relations (\ref{eq:pair-off_omega}) and the following symmetries:
\begin{equation}
\begin{split}
 &W_{k_1,k_2,k_1,k_2} = W_{k_2,k_1,k_2,k_1} = W_{-k_1,k_2,-k_1,k_2}, \,\,\qquad \\
 &W_{k_1,k_2,k_1,k_2} + W_{\frac{N}{2}-k_1,k_2,\frac{N}{2}-k_1,k_2} -W_{-k_1,k_2,-k_1,k_2} 
 - W_{-\frac{N}{2}+k_1,k_2,-\frac{N}{2}+k_1,k_2} = 0\,,
 \end{split}
 \end{equation}
valid for all admissible values of $k_1, k_2$.
It can be proven that $N-1$ functionally independent invariants 
exist and are in involution (for $\alpha$ and $\beta$-FPUT see the proof of integrability in \cite{henrici2008results,rink2006proof}).  We present these invariants in a way that highlights the fact that the resonant quartets are disconnected:

For each $k = 1, \ldots, \lfloor N/4\rfloor$, define an irreducible quartet as the set $\displaystyle Q_k = \left\{b_k, b_{-k}, b_{\frac{N}{2}-k}, b_{-\frac{N}{2}+k}\right\}$. There are four different modes in $Q_k$, except for the degenerate case $k=N/4$, valid when $N/4$ is integer, where $Q_{N/4}$ has two different modes. In the non-degenerate case, the following four invariants depend on the four modes in $Q_k$:
\begin{itemize}
\item 3 quadratic invariants: 
$$|b_k|^2 + |b_{-k}|^2, \qquad |b_k|^2 + |b_{-\frac{N}{2}+k}|^2 , \qquad |b_{\frac{N}{2}-k}|^2 + |b_{-\frac{N}{2}+k}|^2. $$
\item 1 quartic invariant:
$$2 W_{k,\frac{N}{2}-k,-k,-\frac{N}{2}+k} \left(b_k^* b_{\frac{N}{2}-k}^* b_{-k} b_{-\frac{N}{2}+k} + c.c.\right) + W_{k,k,k,k} |b_k|^2 |b_{-k}|^2 +  W_{\frac{N}{2}-k,\frac{N}{2}-k,\frac{N}{2}-k,\frac{N}{2}-k} |b_{\frac{N}{2}-k}|^2 |b_{-\frac{N}{2}+k}|^2.$$
\end{itemize}

In the degenerate case, valid when $N/4$ is integer, we have $Q_{N/4}=\left\{b_{\frac{N}{4}}, b_{-\frac{N}{4}}\right\}$ and the following two invariants depend on the two modes in $Q_{N/4}$:
\begin{itemize}
\item 1 quadratic invariant:
$$|b_{\frac{N}{4}}|^2 + |b_{-\frac{N}{4}}|^2\,.$$
\item 1 quartic invariant:
$$W_{\frac{N}{4},\frac{N}{4},-\frac{N}{4},-\frac{N}{4}} \left[\left(b_{\frac{N}{4}}^* b_{-{\frac{N}{4}}}\right)^2 + c.c.\right] + W_{\frac{N}{4},\frac{N}{4},\frac{N}{4},\frac{N}{4}} |b_{\frac{N}{4}}|^2 |b_{-\frac{N}{4}}|^2 .$$
\end{itemize}

Thus, in terms of counting, any irreducible quartet (degenerate or not) contributes with a number of invariants that is equal to the number of modes in the quartet. Finally, notice that the mode $b_{\frac{N}{2}}$ is not in any irreducible quartet. In fact, the following is an invariant: $|b_{\frac{N}{2}}|^2$.

It is thus easy to show by simple counting of the modes in the irreducible quartets that the system has a total of $N-1$ functionally independent invariants:
\begin{enumerate}
\item When $N/4$ is not integer, the irreducible quartets give a total of $4 \times \lfloor N/4 \rfloor = N - 2$ invariants. The missing invariant is $|b_{\frac{N}{2}}|^2$.
\item When $N/4$ is integer, we get a total of $4 \times (N/4 -1)= N -4$ invariants from the non-degenerate quartets, plus $2$ invariants from the degenerate quartet $Q_{N/4}$, totalling  again $N - 2$ invariants. The missing invariant is $|b_{\frac{N}{2}}|^2$.
\end{enumerate}

The result proofs the asymptotic integrability of the DNKG model up to four wave interactions and justifies the 
metastable states observed in numerical simulations previously performed, \cite{fucito1982approach}.



\subsection{Higher order resonances and break-down of integrability}

For the discussion above, we see that the 3- and 4-wave collision terms cannot be effective in bringing the system to equipartition. The isolated resonant quartets of the 4-wave integrable Hamiltonian do not bring the system in a thermalized state and resonances at higher order are to be investigated.
For all discrete models considered here we have to perform an extra canonical change of variables, from the normal modes $b_k$ to some other variables $c_k$, such that in the new variables the nonresonant terms are removed from the Hamiltonian.
The question is now what is the leading order resonant wave interaction.
In \cite{Onorato2015}, only power-of-two values of $N$ were investigated (akin to the original paper on the $\alpha$-FPUT problem) and five-wave interactions were excluded on numerical grounds. From a more recent investigation \cite{bustamante2018exact}, it turned out that five-wave resonant interactions exist only if $N$ is divisible by 3 when $m=0$.
When $N = 2^a 3^b$ with $a, b >1$, the resulting five-wave clusters are connected and, in principle, a thermalized state can be reached, but this requires further study, to be discussed in a subsequent paper.

Excluding such specific values of $N$, six-wave resonant interactions are the leading order processes: it is always possible to find resonant six-wave tuples that are all interconnected and cover the whole Fourier space.
These resonances are due to the same symmetries of $\omega_k$ and are of the form:
\begin{equation}
\label{eq:umklapp61}
k_1+k_2+k_3=-k_1-k_2-k_3\mod N,\,k_1+k_2+k_3=0\mod N.
\end{equation}
Such resonances are valid for even and odd values of  $N$. Additional resonances can be found both in pairing-off form for even $N$, and recently also other resonances not in pairing-off form have been found \cite{bustamante2018exact}, but they are not necessarily to cover the whole Fourier space and we will not discuss them.
Another six-wave processes is theoretically possible, that is $4\rightarrow 2$. Explicit formulas for $4\rightarrow 2$ resonances have been found recently in\cite{bustamante2018exact}, at least for $N$ divisible by 3. 

Excluding those specific values of $N$ for which five wave interactions and  $4\rightarrow 2$ (or $4\rightarrow 2$) exist, the six wave interaction Hamiltonian for the $\alpha$, $\beta$-FPUT and DNKG equation can be written as 
\begin{equation}
\label{eq:Hnlin6}
\frac{H}{N}= \frac{H_{\text{integrable}}}{N}+\sum_{1,2,3,4,5,6} Z_{1,2,3,4,5,6}
c_1c_2c_3c_4^*c_5^*c_6^*\delta^{(N}_{1+2+3-4-5-6},
\end{equation}
where $\delta_N$ is the Kroneker modulo $N$.
The new matrix interaction can be calculated~\cite{laurie2012one}, but its precise form is not relevant to our scope. 

\subsection{The time scale of thermalization}
The next step to be taken in order to evaluate the thermalisation time-scale should be 
the derivation of  the corresponding kinetic equation, as done in the continuous case.
Unfortunately, the kinetic equation can be rigorously derived only in the thermodynamic limit, that is $N\rightarrow\infty$. 
Deriving a \textit{discrete} version ($N$ finite) of the kinetic equation poses significant mathematical problems\cite{eyink2012kinetic}.

Here, we make the conjecture that the time scale for thermalization in 
the discrete dynamics corresponding to the 6-wave interaction given by eq. (\ref{eq:Hnlin6}) is at the leading order equivalent to the corresponding continuous 6-wave process.  We also assume that the integrable part of the dynamics does not 
lead to any irreversible transfer of energy and the irreversible dynamics is fully contained in the six-wave interaction term. 
Our conjecture is therefore tantamount to saying that the thermalisation time-scale in the discrete case will be always proportional to the square of the coupling coefficient $Z_{1,2,3,4,5,6}$ in the six  by Eq. (\ref{eq:Hnlin6}), the latter being proportional to $\alpha^{4}$ in the $\alpha$-FPUT model and  $\beta^{2}$ in the  $\beta$-FPUT and DNKG system.
We get the following estimates for the thermalisation time-scale
for the $\alpha$-FPUT model
\begin{equation}
\label{eq:gammaalpha}
T_\text{eq}\propto\alpha^{-8}\propto \epsilon_{\alpha}^{-4}
\end{equation}
and 
\begin{equation}
\label{eq:gammabeta}
T_\text{eq}\propto\beta^{-4}\propto \epsilon_{\beta,KG}^{-4}
\end{equation}
for the $\beta$-FPUT, DNKG models.
The argument used in the discrete case is always made for the limit of vanishing nonlinearity, where only exact resonances are important.
However, the presence of a small but finite nonlinearity is expected  to broaden the frequencies.
In the case of a small number of modes $N$, that is in the present discrete case, an important difference with respect to the continuous case is expected. 
Since the wave-space is discrete, a large enough nonlinearity may cause a sufficient broadening to trigger resonances which are not in the exact-resonance manifolds and are found at a lower-order than the exact ones. 
Therefore, we can expect that at small $N$ and sufficient nonlinearity the transfer of energy can be made through lower-order processes on smaller time-scales. A definite answer on the role of quasi-resonance is
beyond the scope
    of the present manuscript.

\subsection{Final remarks on the Wave Turbulent approach}
Concluding the theoretical analysis, the WT formalism applied to the different anharmonic chains leads to consider the problem of transfer of energy as related to the collisions between waves in the system. In this framework, the energy is redistributed among modes if resonances exist which connect the whole wave-space.
In the thermodynamic limit, the theory is continuous and is almost rigorous. It predicts that 4-wave processes are the leading order process in all chains and this fact allows a transparent estimate of the equipartition time-scale.
In the case of a finite small $N$, we have considered the system intrinsically discrete and we have searched for the exact discrete resonances. We have seen that in this case, 4-waves processes are not able to transfer energy among modes, nor 5-wave collisions. The leading order is found to be 6-waves. 
Although we are not able to write the corresponding discrete kinetic equation, we have made the hypothesis that the same reasoning  as in the continuous case can be followed, at least when looking at statistical observables like equipartion time-scale.
In this way, an estimate of this time-scale is obtained always proportional to the square of the nonlinear coupling coefficient.
Moreover, also when $N$ is small, nonlinear effects can be important.
We expect that the broadening of the frequencies due to nonlinearity can trigger lower-order process to transfer energy among modes when it becomes 
of the same order of the spacing in the frequency space, which is inversely proportional to $N$.
The discrete case stands therefore on a less rigorous ground and the conjectures proposed can be only verified numerically \emph{a posteriori}.


\section{Numerical simulations}

We now present the result of numerical simulations in support of the previous discussion. The equations (\ref{eq:eqm_a}) and (\ref{eq:eqm_b}) have been implemented in physical space, with a symplectic algorithm of the sixth order\cite{yoshida1990construction}. It is important to use a high-order integration scheme that preserves accurately  the energy of the system, because simulation times are very long compared to the typical wave periods. The initial random  energy per mode, $e_k$, were drawn from an uniform distribution, and then scaled in order to obtain the desired total linear energy $E$ and an initial number of particles $\mathcal{N}$ compatible with a final relaxation distribution with $\mu=0$ in eq.~(\ref{eq:RJ}), in order to better observe equipartition.
\begin{figure}
\centering
\includegraphics[width=0.5\textwidth]{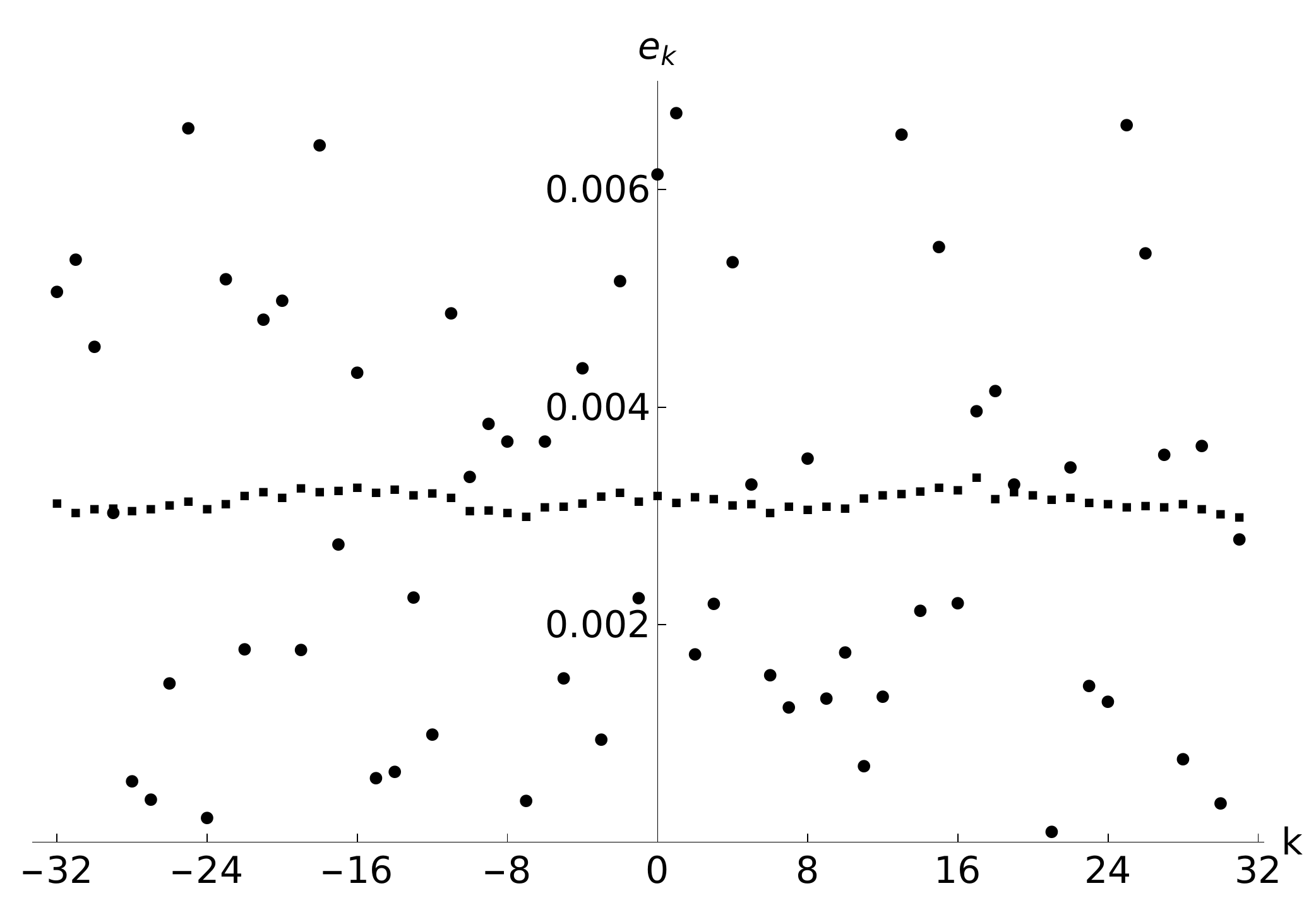}
\caption{The initial distribution of $e_k$ for the DNKG model with $m=1$, $N=64$ (circles), with $E=0.2$ and $\mathcal{N}\simeq 0.129$. The thermalized final state is shown with the squares, and it approximately corresponds to a Rayleigh-Jeans distribution~(\ref{eq:RJ}) with $\mu=0$.}
\label{fig:initial} 
\end{figure}
As an example, we show the initial distribution of $e_k$ for the simulations of the DNKG system with $N=64$ in figure~\ref{fig:initial}, together with its final thermalized state. The linear energy $E$ and the wave action $\mathcal{N}$ change only for one part in $10^4$ across the whole simulation time, being quasi-conserved quantities.
Each realization in the ensemble shares the same initial energy per mode, but phases are randomized. This scheme of initialization ensures that all the realization share the same initial linear energy. The time step was set to $0.1$ in all simulations, and it was checked that from beginning to the end of the simulation the total energy is conserved up to one part in $10^7$. The computation is very easy to parallelize, since ensemble averages need to be computed. 
For this reason we implemented the code on GPU hardware, which is powerful when the code can be parallelized. 

The parameter $\alpha$ and $\beta$ are chosen in such a way that the nonlinear energy is small compared to the linear one
and the parameters $\epsilon_{\alpha}$ and  $\epsilon_{\beta,KG}$ are small.
We will present the results of $T_\text{eq}$ as a function of $\epsilon_{\alpha}$ and  $\epsilon_{\beta,KG}$.

The approach to equipartition is often monitored with the information entropy,
\begin{equation}
\label{eq:Sinf}
S_\text{inf}=\sum_{k=0}^{N-1}e'_k\log(e'_k),\,e'_k=\frac{N}{E}e_k,
\end{equation}
where 
\begin{equation}
e_k=\omega_k \langle |a_k|^2 \rangle\;\;\; \text{and}\;\;\; E=H_{\text{lin}}=\sum_{k=0}^{N-1}e_k.
\end{equation}
In general, the information entropy is not guaranteed to have entropy-like properties, that is monotonic increase or decrease over time. Nevertheless, it has the useful property that at perfect equipartition $S_\text{inf} = 0$  and it is greater than zero in any other state. We argue  that when $\mu=0$ the information entropy is essentially equivalent to the WT entropy~(\ref{eq:entropy}). The WT entropy has the remarkable property that one can replace $n_k$ in eq. (\ref{eq:entropy}) with $f(k)n_k$, with $f(k)$ being any function of $k$ constant in time and still it maintains the property $\mathrm{d}S/\mathrm{d}t\geq 0$. We exploit this and use $f(k)=N/E\omega_k$, so that the WT entropy now is computed as
\begin{equation}
\label{eq:Swt}
S'_\text{WT}=-\sum_{k=0}^{N-1}\log(e'_k).
\end{equation}
This expression preserves the monotonicity of the WT entropy with an inverted sign ($\mathrm{d}S'_\text{WT}/\mathrm{d}t\leq0$, $S'_\text{WT}=0$ at equipartition). Since $\log(e'_k)\simeq 0$ close to equipartition, we see that in practice the two entropies~(\ref{eq:Sinf}) and~(\ref{eq:Swt}) are equivalent for the purpose of determining the equipartition of the system. 
In numerical simulations, the entropy shows some fluctuations and never reaches exactly zero, because the simulation time is finite, but most importantly because the ensemble size is finite. In the following, we will determine $T_\text{eq}$ looking at the time when the entropy crosses a threshold value from above. The selection of the threshold depends on the ensemble size and the number of particles, but it does not depend on the degree of nonlinearity. Generally we chose a value of $S'_\text{WT}$ two orders of magnitude smaller than  that of the initial conditions.

\subsection{Small number of particles, discrete resonances}
\begin{figure}
\centering
\includegraphics[width=8cm]{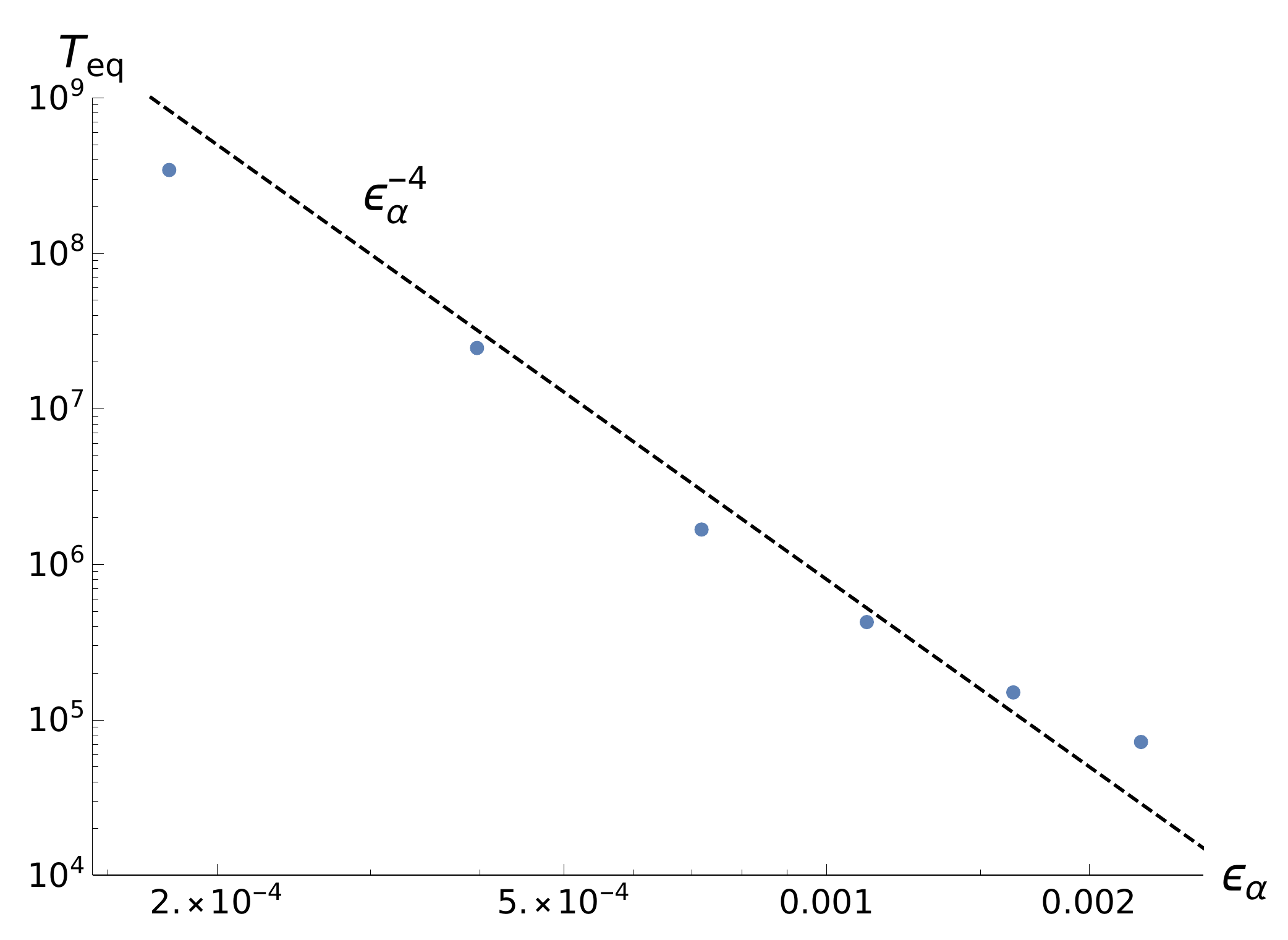}
\caption{$\alpha$-FPUT, $N$=64, $E=1$, ensemble size 4096.}
\label{fig:alpha64}
\end{figure}
\begin{figure}
\includegraphics[width=8cm]{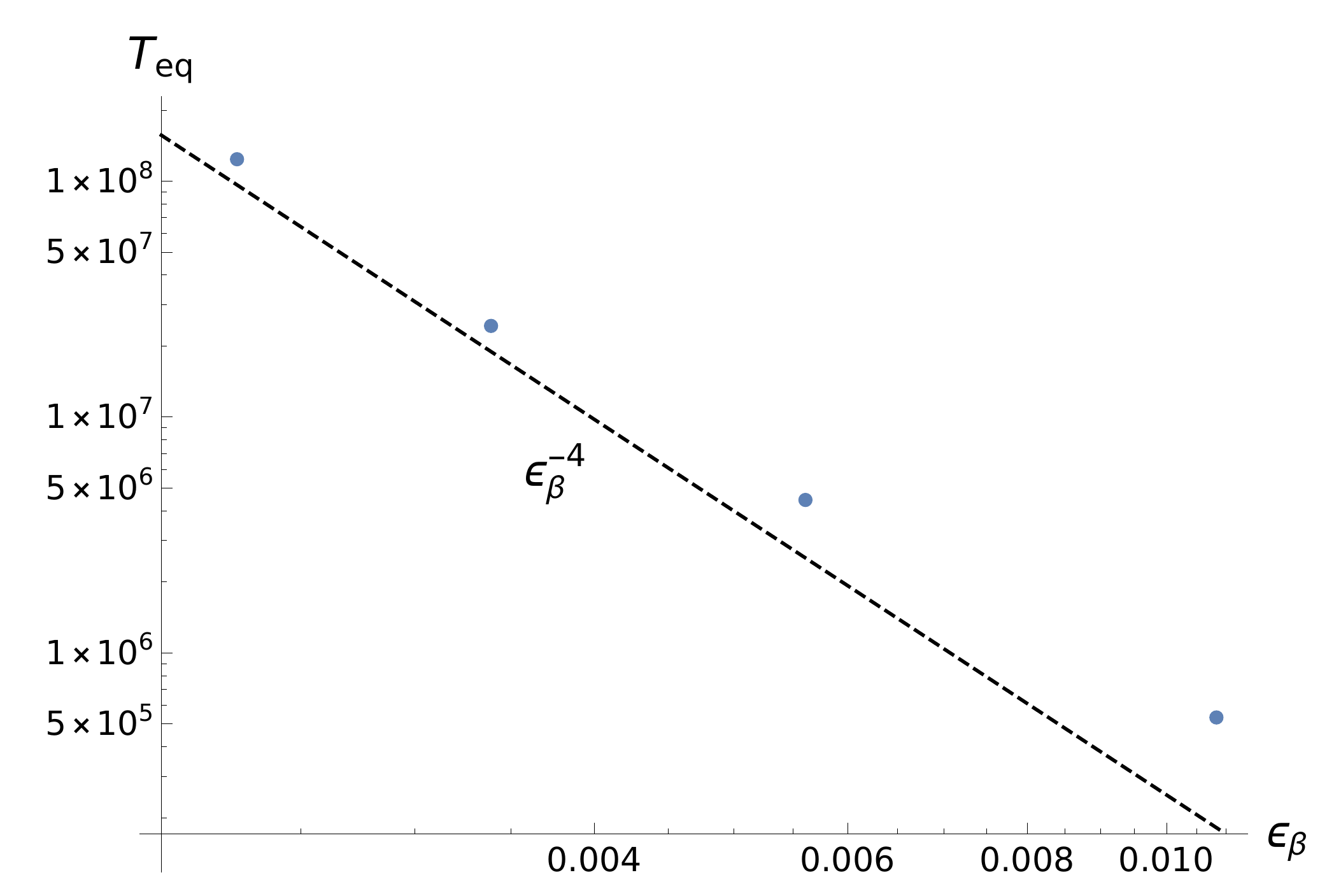}
\caption{$\beta$-FPUT, $N$=64, $E=1$, ensemble size 4096.}
\label{fig:beta64}
\end{figure}
\begin{figure}
\includegraphics[width=8cm]{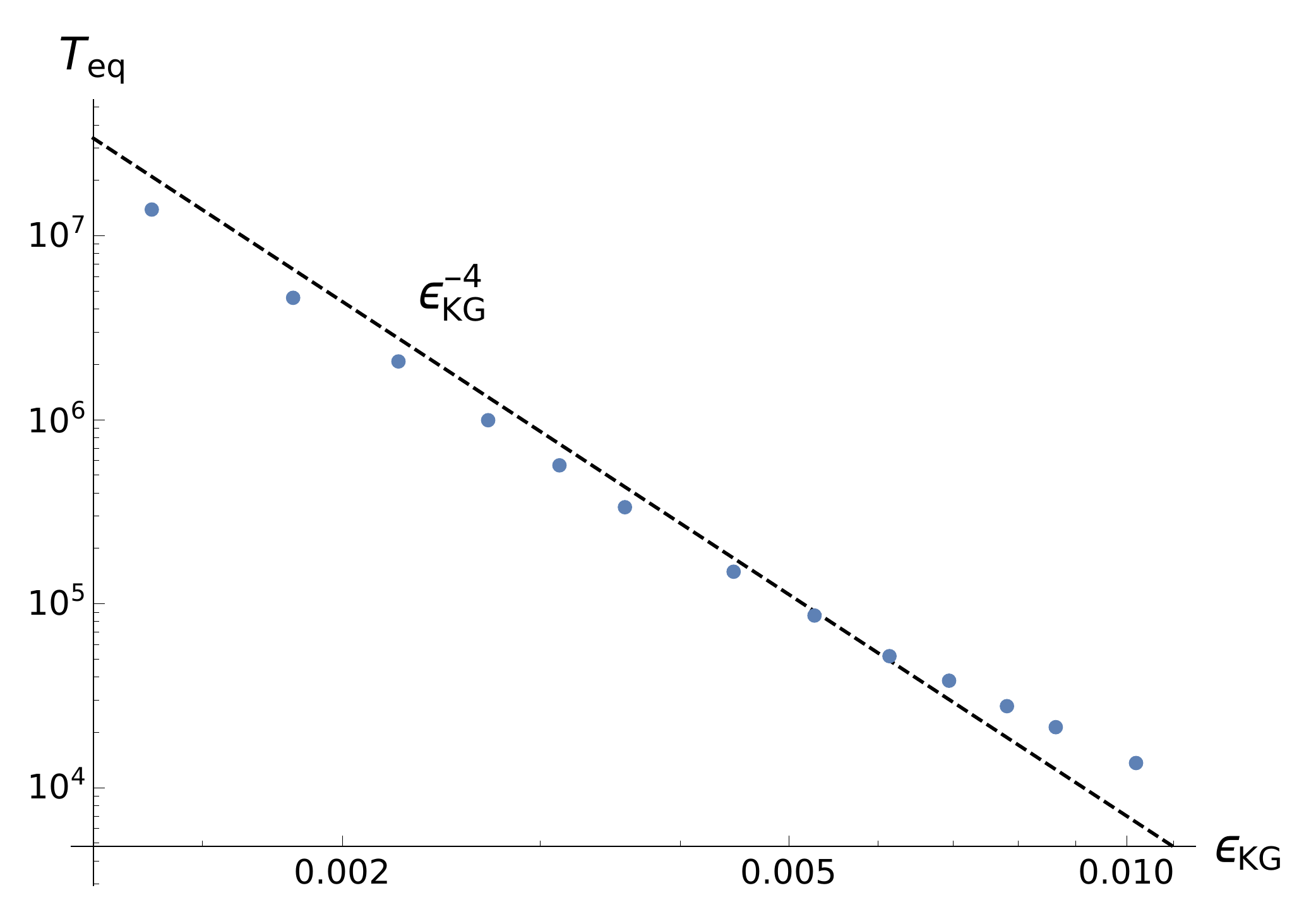}
\caption{DNKG, $N$=64, $m=1$, $E=0.2$, ensemble size 4096.}
\label{fig:kg64}
\end{figure}
We first present the results on the case of a limited number of particles, $N=64$, so that discrete resonances need to be considered. To recap, we expect for the $\alpha$-FPUT system the scaling
\begin{equation}
T_\text{eq}\propto\alpha^{-8}\propto\epsilon_{\alpha}^{-4}
\end{equation}
and for $\beta$-FPUT and DNKG systems the scaling
\begin{equation}
T_\text{eq}\propto\beta^{-4}\propto\epsilon_{\beta,KG}^{-4}.
\end{equation}
We show the results of the simulations for $N=64$ in figures \ref{fig:alpha64}, \ref{fig:beta64}, \ref{fig:kg64}. We see a good alignment with the expected power-laws for all the three models. We chose $m=1$ in the DNKG model arbitrarily: in\cite{pistone2018thermalization} it is  argued that since the resonances in the discrete case do not depend on $m$, then in the scope of the thermalization dynamics the value of $m$ is not relevant.
However, we should add here a comment for the limit of $m$ very small. In such cases, the contribute to $H_\text{nlin}$ increases for the low frequency modes. 
On the other hand, for the FPUT systems the nonlinearity depends on a power of the finite differences, which are approximately proportional to $k$,
hence the contribute to the nonlinear part of the Hamiltonians for modes with small $k$ does not grow.
\begin{figure}
\centering
\includegraphics[width=0.5\textwidth]{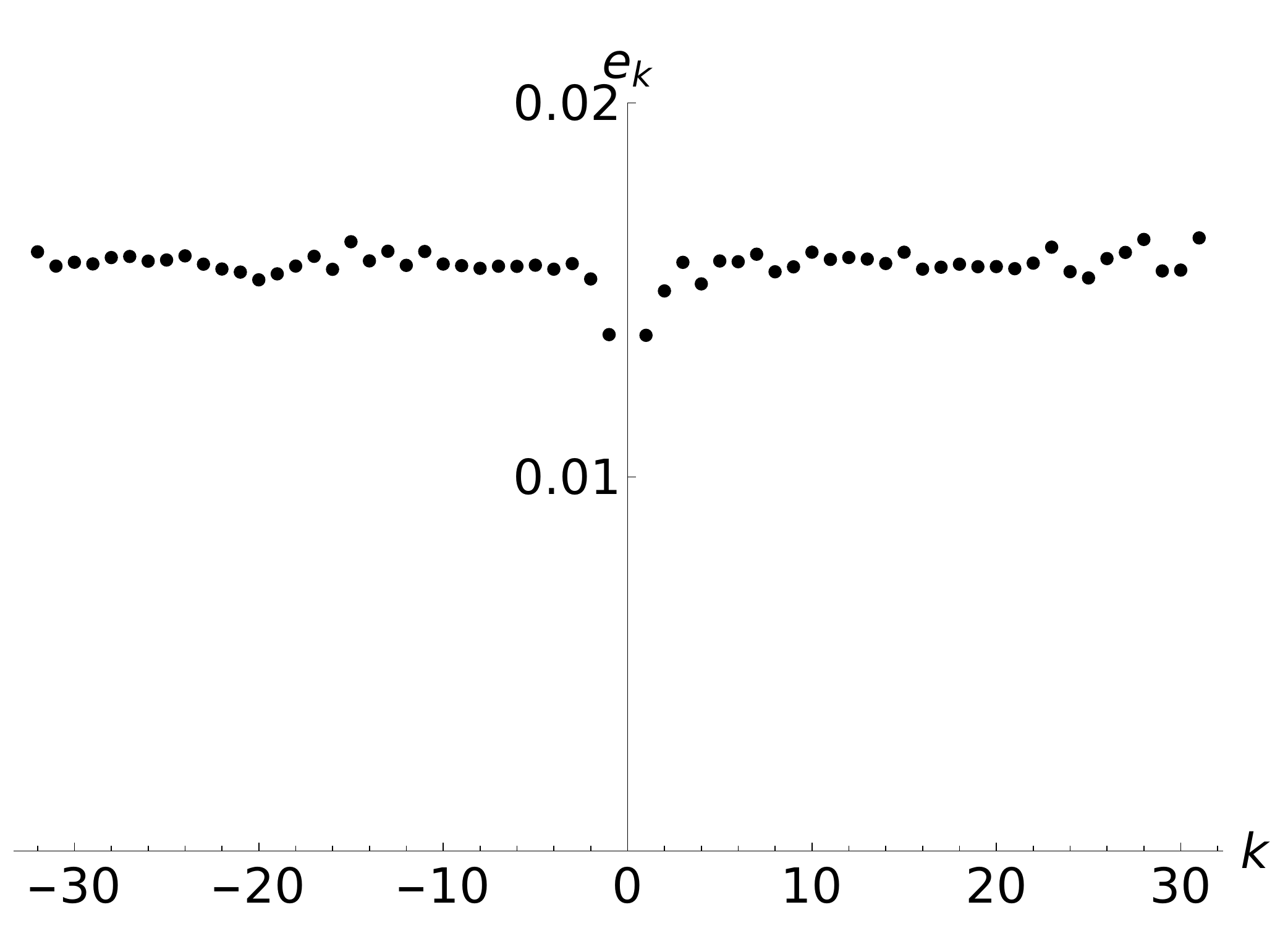}
\caption{The equilibrium linear energy of the DNKG model with $N=64$, $m=0$, $E=1$. The modes close to $k=0$ do not thermalize as the other modes, because their contribute to $H_\text{nlin}$ is significantly higher than the other modes.}
\label{fig:KGm0} 
\end{figure}
As a consequence, in the case of the DNKG system with $m=0$, a complete thermalization of the lower modes would invalid the weak-nonlinear assumption, because the nonlinearity would be too high. What is observed actually (see figure \ref{fig:KGm0}) is a partial thermalization of the system, with the lower modes stationary at a lower energy than equipartition.

\subsection{Large number of particles, continuous resonances}

\begin{figure}
\includegraphics[width=8cm]{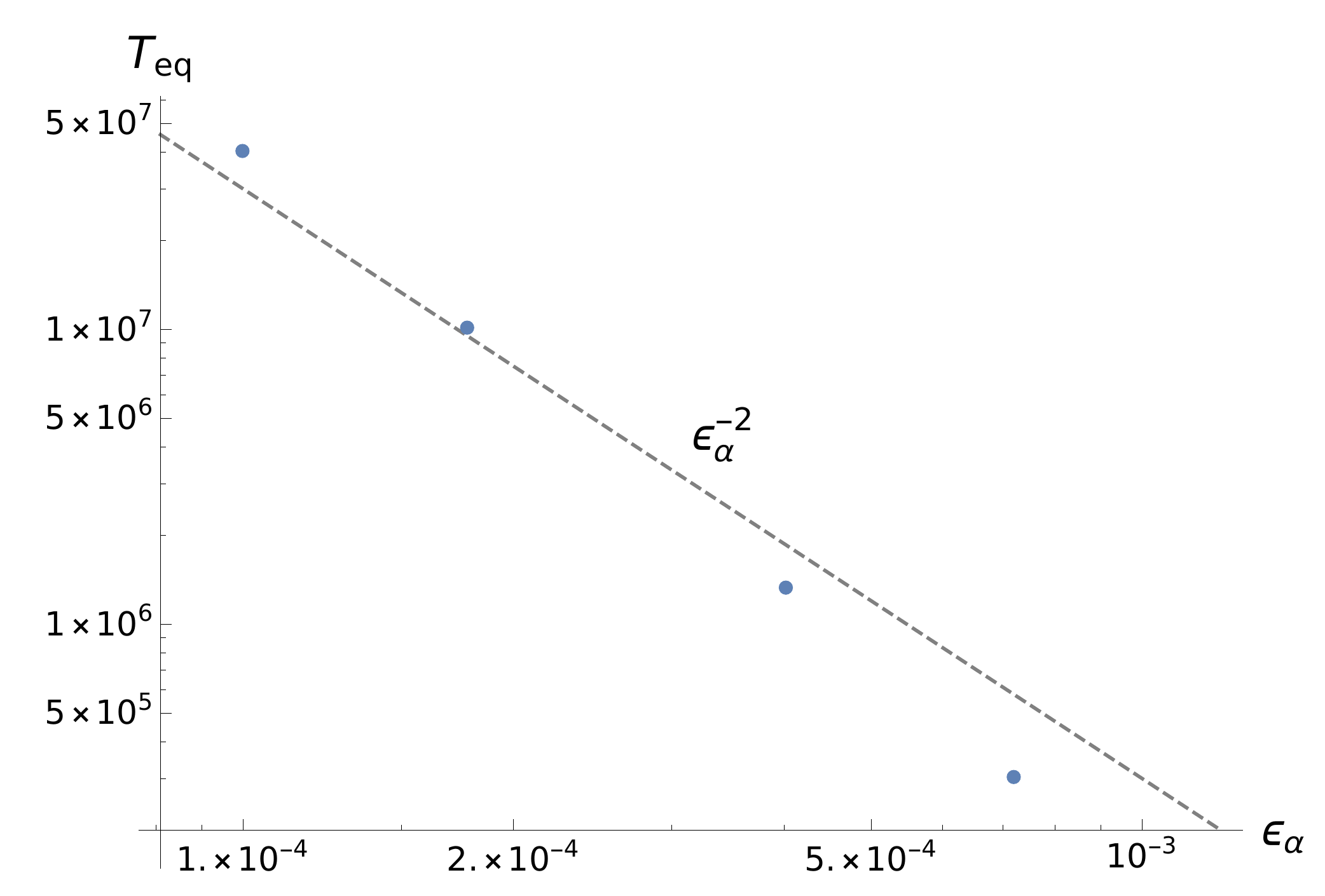}
\caption{$\alpha$-FPUT, $N=1024$, $E=16$, ensemble size 4096.}
\label{fig:alpha1024}
\end{figure}

\begin{figure}
\includegraphics[width=8cm]{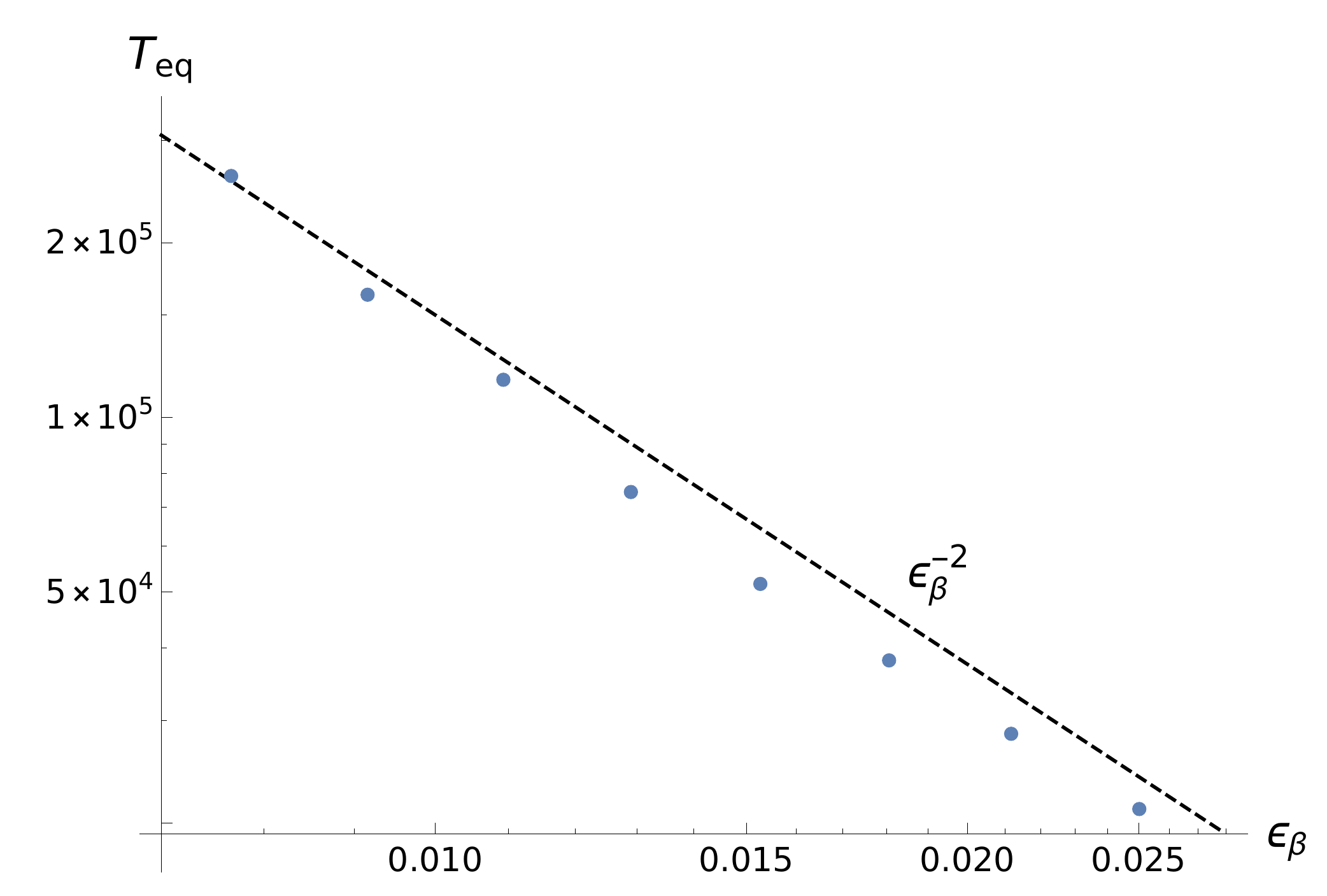}
\caption{$\beta$-FPUT, $N=1024$, $E=16$, ensemble size 1024.}
\label{fig:beta1024}
\end{figure}

\begin{figure}
\includegraphics[width=8cm]{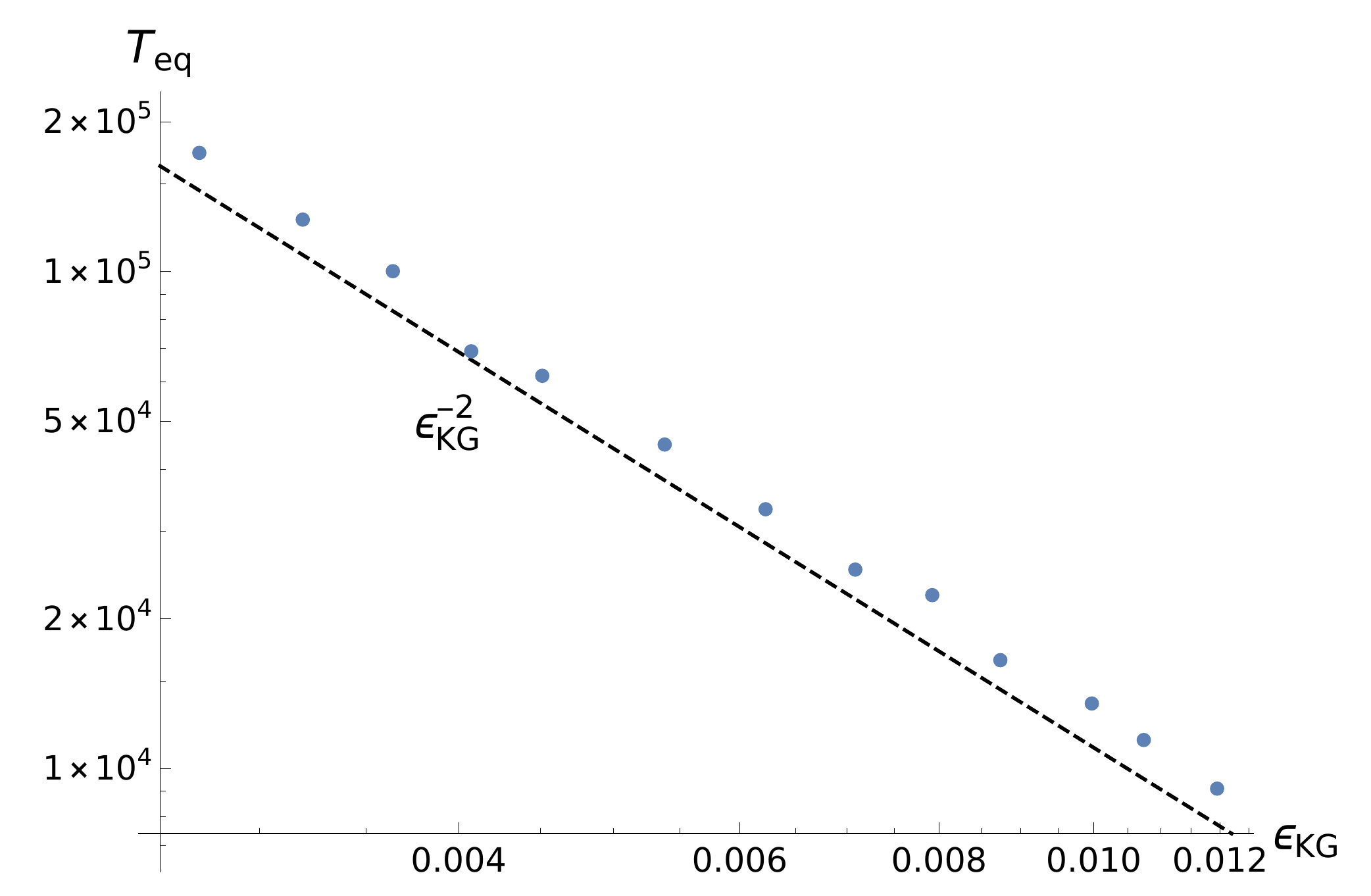}
\caption{DNKG, $N=1024$, $m=1$, $E=3.2$, ensemble size 4096.}
\label{fig:kg1024}
\end{figure}
We now present the results of numerical simulations for the continuous resonance manifold. The expected results are
\begin{equation}
T_\text{eq}\propto\alpha^{-4}\propto\epsilon_{\alpha}^{-2}
\end{equation}
for the $\alpha$-FPUT system, and
\begin{equation}
T_\text{eq}\propto\beta^{-2}\propto\epsilon_{\beta,KG}^{-2}
\end{equation}
for $\beta$-FPUT and DNKG systems. The number of particles and the strength of the nonlinearity necessary to observe the continuous resonances are not something predicted by our treatment. 
However, estimates of the frequency broadening are  available in\cite{lvov2018double,nazarenko2011wave}, and they can be compared to the typical frequency spacing for a fixed number of particles. We point out that in general a higher value for $m$ in eq.~(\ref{eq:omega}) makes the frequency gaps smaller among modes, hence the DNKG model in general requires a small number of particles to observe the continuous resonances.

In figure \ref{fig:alpha1024},\ref{fig:beta1024}, \ref{fig:kg1024} we show the results for simulations with $N=1024$, again with $m=1$ for the DNKG system. The figure 
highlight a reasonable agreement between numerics and   the theoretical predictions (the exponent in the $\alpha$-FPUT seems to be slightly steeper than predicted, probably because the thermodynamic limit has not been reached numerically yet).
Therefore, from the simulation data we can deduce that the WT picture works reasonably well in all cases, at small and large $N$ .

\section{Conclusions}

In this work, we have presented recent results on the dynamics of nonlinear chains, focusing on the thermalisation time.
We have proposed to encompass the behaviour of all systems of this kind within the Wave Turbulence framework. 
In this context, the universal mechanism invoked to lead the systems to equipartition is the resonance among different modes, at least when generic initial conditions are considered.
The scaling of the thermalization time $T_\text{eq}$ as a function of the nonlinearity level of the system is obtained from the theory.
The result is a power-law for all the cases considered, with an exponent dependent on the order of the active resonances in the system.
Although the point if view is rather different, present results are not in contradiction with the most recent and accurate estimates obtained also for particular low-wavenumber initial conditions~\cite{benettin2011time}.
Since the resonant manifold is significantly different in the discrete and continuous $N\rightarrow\infty$ limit, we have considered the two cases separately, and they lead to two different scalings.
Notably, large-size chains reach equipartition faster.

Our aim in this paper has been to present the results anticipated in  works\cite{Onorato2015,lvov2018double,pistone2018thermalization} in an unified manner to underline how our approach can be systematic, rather than dependent on the particular features of the systems. 
In the continuous limit, we have  shown that  there exists a resonant manifold of $2\rightarrow 2$ waves. 
We have analysed numerically the continuous limit $N\rightarrow\infty$ for the $\alpha$ and $\beta$-FPUT systems, as done previously for the DNKG chain. In the discrete case, for $N$ a power of two (which excludes the existence of five-wave resonant interactions  \cite{bustamante2018exact}),  we have extensively verified that, for all the considered systems, the leading order interaction consists of six wave interactions. 
All the results seem to indicate a universal route to thermalization predicted. We have also shown that, besides the  $\alpha$ and $\beta$-FPUT systems, also the DNKG equation is asymptotically integrable, if the expansion is truncated at four wave interactions.

One should note that the WT theory predicts thermalization  for arbitrary small nonlinearity, with always the same power law scaling. It would be interesting to understand 
how this is related to the result of Nekhoroshev {\cite{nekhoroshev1977exponential} who demonstrated that
solutions stay close to their integrable counterparts  for an exponentially long time.



\section*{Appendix}
Let's define the function $f_k=2 |\sin(k/2)|$ as  the linear dispersion relation for  the $\alpha$ and $\beta$- FPUT models
which corresponds to the general dispersion relation, $\omega_k$, when $m=0$.
In \cite{bustamante2018exact} it has been shown that all processes $X\rightarrow$1 are forbidden for the dispersion 
relation $f_k$.
\begin{equation}
f_{k_1}+f_{k_2}+...+f_{k_X}\ge f_{k_1+k_2+...+k_X},
\label{subadd}
\end{equation}
where the equality holds only for wave numbers equal to 2$l\pi$, with $l\in{\mathbb{Z}}$. Here we show that
\begin{equation}
\omega_{k_1}+\omega_{k_2}+...+\omega_{k_X}> \omega_{k_1+k_2+...+k_X}
\label{eq:thesis}
\end{equation}
for any value of $m>$0.
In order to show it, we square (\ref{eq:thesis}) and, after re-arranging, we get:
\begin{equation}
F(k_1,k_2,..,k_X;m)=(X-2) m + f_{k_1}^2+f_{k_2}^2+..+f_{k_X}^2-f_{k_1+k_2+..+k_X}^2+
\sum_{i=1}^{X}\omega_{k_i}\sum_{j=1}^{X}\omega_{k_j}-\sum_{j=1}^{X}\omega_{k_j}^2>0
\end{equation}
The function $f_k$ is $m$-independent, while $\omega_k$ with $m\ne0$ is proportional to $m$.
The terms of the type $-\omega_{k_i}^2$ all cancel out. 
It is easy to observe that $F(k_1,k_2,..,k_X;m)$ is an increasing monotnic function of $m$.
For $m=0$, because of the inequality in (\ref{subadd}), the $F(k_1,k_2,..,k_X;0)\ge 0$;
then, for its monotonicity, for any $m>0$, $F(k_1,k_2,..,k_X;0)> 0$. This proves that there 
are no resonances in the DNKG model of the type $X\rightarrow$ 1.

\section*{Acknowledgments}
M. O. has been funded by Progetto di Ricerca d'Ateneo CSTO160004. M.O. was supported by the ``Departments of Excellence 2018-2022'' Grant awarded by the Italian Ministry of Education, University and Research (MIUR) (L.232/2016).
Simulations were run on GPU hardware provided at the OCCAM facility, University of Turin.

\bibliography{references}{}
\bibliographystyle{plain}

\end{document}